\documentclass[aps,pra,reprint,amsmath,amssymb]{revtex4-2}
\usepackage{lipsum} 
\usepackage{ragged2e} 

\usepackage{graphicx}
\usepackage{dcolumn}
\usepackage{bm}
\usepackage{float}
\usepackage[per-mode=power, 
            exponent-product = \times, 
            inter-unit-product = \ensuremath{{\cdot}}, 
            tight-spacing = true,
            number-unit-product=~,
            parse-numbers=false,]{siunitx}
\AtBeginDocument{\RenewCommandCopy\qty\SI}
\ExplSyntaxOn
\msg_redirect_name:nnn { siunitx } { physics-pkg } { none }
\ExplSyntaxOff

\newcommand{\SiN}{Si\textsubscript{3}N\textsubscript{4}}
\usepackage{siunitx}
\usepackage{soul}
\newcommand{\um}[1]{\SI{#1}{\micro\meter}} 

\begin{document}

\title{Highly squeezed nanophotonic quantum microcombs\\
with broadband frequency tunability}

\author{Yichen Shen$^{1\dagger}$}
\author{Ping-Yen Hsieh$^{2,1\dagger}$}
\author{Dhruv Srinivasan$^{1}$}
\author{Antoine Henry$^{1}$}
\author{Gregory Moille$^{3,4}$}
\author{Sashank Kaushik Sridhar$^{1}$}
\author{Alessandro Restelli$^{3}$}
\author{You-Chia Chang$^{2}$}
\author{Kartik Srinivasan$^{3,4}$}
\author{Thomas A. Smith$^{5}$}
\author{Avik Dutt$^{1,6}$}
\thanks{Corresponding author: avikdutt@umd.edu}

\affiliation{$^{1}$Department of Mechanical Engineering and Institute for Physical Science \& Technology, University of Maryland, College Park, MD 20742, USA}
\affiliation{$^{2}$Department of Photonics, College of Electrical and Computer Engineering, National Yang Ming Chiao Tung University, Hsinchu City 30069, Taiwan}
\affiliation{$^{3}$Joint Quantum Institute, National Institute of Standards and Technology, University of Maryland, College Park, MD 20742, USA}
\affiliation{$^{4}$Microsystems and Nanotechnology Division, National Institute of Standards and Technology, Gaithersburg, MD 20899, USA}
\affiliation{$^{5}$Quantum Research and Applications Branch, Naval Air Warfare Center Aircraft Division, Patuxent River, MD 20670, USA}
\affiliation{$^{6}$National Quantum Laboratory (QLab) at Maryland, College Park, MD 20740, USA}

\thanks{These authors contributed equally to this work.}

\date{\today}
\begin{abstract}
Squeezed light offers genuine quantum advantage in enhanced sensing and quantum computation; yet the level of squeezing or quantum noise reduction generated from nanophotonic chips has been limited. In addition to strong quantum noise reduction, key desiderata for such a nanophotonic squeezer include frequency agility or tunability over a broad frequency range, and simultaneous operation in many distinct, well-defined quantum modes (qumodes). Here we present a strongly overcoupled silicon nitride squeezer based on a below-threshold optical parametric amplifier (OPA) that produces directly detected squeezing of 5.6 dB $\pm$ 0.2 dB, surpassing previous demonstrations in both continuous-wave and pulsed regimes. We introduce a seed-assisted detection technique into such nanophotonic squeezers that reveals a quantum frequency comb (QFC) of 16 qumodes, with a separation of 11~THz between the furthest qumode pair, while maintaining a strong squeezing. Additionally, we report spectral tuning of a qumode comb pair over one free-spectral range of the OPA, thus bridging the spacing between the discrete modes of the QFC. Our results significantly advance both the generation and detection of nanophotonic squeezed light in a broadband and multimode platform, establishing a scalable, chip-integrated path for compact quantum sensors and continuous-variable quantum information processing systems.
\end{abstract}

\maketitle

\begin{figure*}[ht]
	\centering
	\includegraphics[width=.9\textwidth]{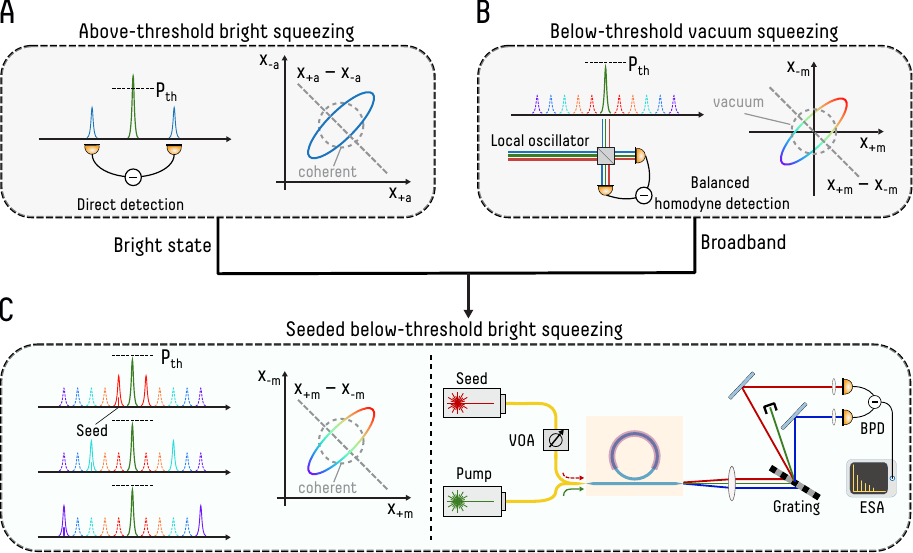} 
	\caption{\textbf{Seed-assisted parametric amplifier as broadband nanophotonic squeezer.}
		(\textbf{A}) When the pump laser power in a microresonator exceeds the oscillation threshold, the $\pm a$ pair of modes with the highest FWM gain is excited. In phase space, this corresponds to a displaced two-mode squeezed state exhibiting squeezing along $x_a-x_{-a}$ quadrature. Because the oscillator is above threshold, these modes carry a mean field amplitude (i.e., they are “bright”) and can be measured via intensity correlations. (\textbf{B}) When the pump power remains below threshold power, multiple signal-idler pairs, denoted as $\pm m$, simultaneously experience parametric gain. This leads to a discrete squeezed frequency comb, a collection of two-mode squeezed states, with each pair showing vacuum squeezing along $x_m-x_{-m}$ quadrature. (\textbf{C}) Introducing a weak seed at one of the $\pm m$ mode stimulates the FWM process, selectively amplifying an idler and generating a correlated signal. In the phase space, each excited mode pair $\pm m$ becomes a displaced two-mode squeezed state, enabling straightforward intensity-difference measurements. Right: simplified setup - a tunable, below-threshold pump laser is combined with a weak seed laser and coupled into the microresonator-based optical parametric amplifier OPA. The generated $\pm m$ signal-idler pairs is spatially separated with a grating and their correlations are detected using a balanced detector (BPD). A variable optical attenuator (VOA) controls the seed power, and the electrical spectrum analyzer (ESA) records the noise spectra.}
	\label{fig1} 
\end{figure*}
\section*{\label{sec:intro}Introduction}
Quantum squeezed states of light, characterized by quantum noise reduction below the shot-noise limit in one quadrature, have emerged as a pivotal resource for quantum sensing \cite{Lawrie:19} and quantum information processing \cite{Braunstein:05}. Increasing the accessible bandwidth and the number of quantum modes, while maintaining strong squeezing, is critical for both the scalability and resource quality of continuous-variable quantum systems \cite{andersen:10}, yet remains a major challenge in nanophotonics. In quantum sensing, strong squeezing directly enhances measurement precision and sensitivity by reducing quantum noise below the standard quantum limit. Broadening the spectral bandwidth and frequency tunability of squeezed light sources significantly expands the measurement rate and utility in spectroscopic sensing \cite{Polzik:1992,herman:25}. In parallel, continuous-variable quantum computing places stringent demands on squeezing strength for fault-tolerant operation \cite{fukui2018high}, while broadband multimode squeezing supports parallelized information encoding and efficient quantum gate implementations \cite{inoue:23}.

Optical parametric oscillators (OPOs) or amplifiers (OPAs) are established nonlinear devices for generating squeezed states. The development of bulk-optic implementations of above-threshold OPOs and below-threshold OPAs have culminated in squeezed light sources generating up to 15~dB of squeezing \cite{wu:86,Vahlbruch:16}, and these devices have been the workhorse of photonic quantum information processing (QIP), enabling the creation of time-multiplexed cluster states \cite{Yokoyama:13,Asavanant:19}, multimode squeezing in wavelength-multiplexed frequency combs \cite{Roslund:14,chen:14,reimer_generation_2016}, and have demonstrated quantum computational advantage \cite{zhong:20,Madsen:22}. In parallel, strong squeezing routinely achieves improved measurement sensitivity, revolutionizing fields ranging from gravitational wave detection \cite{Aasi:13} to molecular spectroscopy and fingerprinting \cite{deAndrade:20,Casacio:21}.

To miniaturize the generation of continuous-variable squeezing across multiple quantum mode (qumode) pairs on compact nanophotonic platforms, Kerr microresonators are increasingly employed. \SiN\,and silica microresonators have produced microcombs supporting a large-scale of discrete quantum modes \cite{Yang:21,Jahanbozorgi:23,wang2025large} and demonstrated multipartite entanglement involving eight modes\,\cite{jia:25}. Typically, measuring these quantum modes requires a phase-coherent local oscillator (LO) \,\cite{Yang:21,tritschler2025,Zhao:20,ulanov2025,wang2025large,Zhang:21,Vaidya:20,Cernansky:20,Chen:22}, a challenging task in nanophotonics where the correlated frequency bins of a two-mode squeezed vacuum can be separated by tens to hundreds of nanometers in wavelength. To circumvent this, inspired by seeding commonly used in off-chip squeezers\,\cite{adamou:25,Li:21,sun:19, mccormick:06}, we leverage stimulated four-wave-mixing (FWM) in a microresonator to generate an OPA-based quantum frequency comb as shown in Fig.\,\ref{fig1}C. A weak seed signal undergoes regenerative gain through optical parametric amplification\,\cite{zhao2023large}, simultaneously generating a correlated idler beam. This process produces bright squeezed beams with nonzero mean amplitudes, enabling direct detection of their correlations via intensity-difference measurement. Our approach bridges the straightforward intensity correlation detection method of above-threshold OPOs (Fig.~\ref{fig1}A) with the broadband multimode capability of below-threshold OPAs (Fig.~\ref{fig1}B).

Using this approach, we demonstrate below-threshold squeezed light with substantial squeezing levels of \qty{5.6}{\dB} $\pm$ \qty{0.2}{\dB}, surpassing prior nanophotonic results in both continuous-wave \cite{shen:25} and pulsed regimes \cite{Nehra:22}. We also demonstrate fine continuous and coarse discrete spectral tunability across one free spectral range (FSR), which indicates a gap-free coverage of the entire quantum comb bandwidth of \qty{11}{\THz}. These advances on the squeezed source side are complemented by a seed-assisted detection strategy, and the introduction of such strategy into nanophotonics provides a practical alternative to standard homodyne detection schemes for harnessing the full extent of quantum resources afforded by microresonators. Unlike recent phase-sensitive amplification (PSA) techniques \cite{Nehra:22,frascella:21,Takanashi:20, shaked:18, virally:21} that employ a secondary OPA stage to amplify quantum states into classical macroscopic states resilient to both noise and loss, our single-stage, seeded OPA produces bright quantum states that mitigate detection noise while remaining sensitive to loss, thereby preserving the quantum nature of the output state. These capabilities pave a versatile path toward integrated photonic quantum systems approaching the challenging requirements of compact squeezers for quantum sensing and QIP.

\section*{Seed-assisted Parametric Amplification for Squeezing}
\begin{figure}
	\centering
	\includegraphics[width=0.45\textwidth]{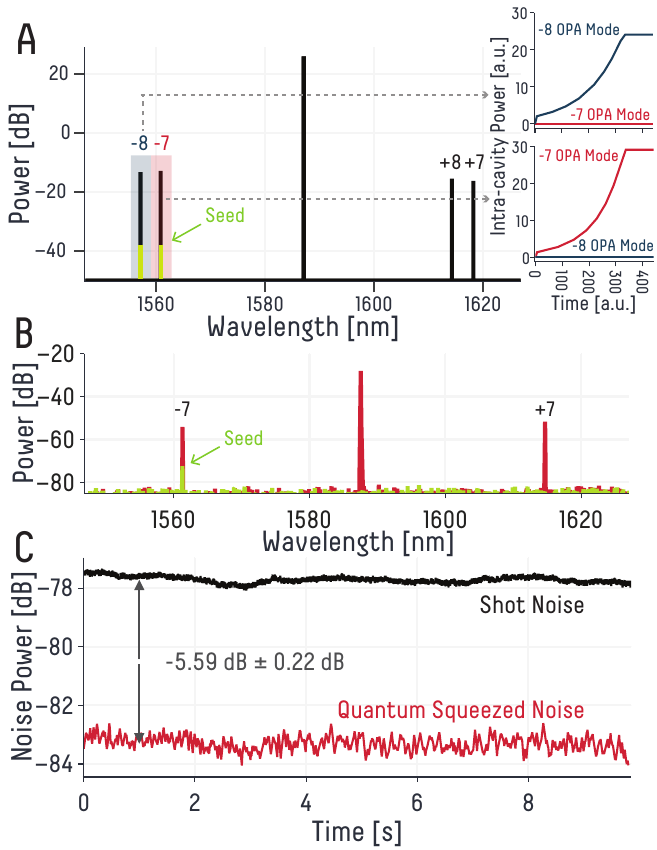} 
	\caption{\textbf{Frequency-multiplexed optical parametric amplifier for quantum squeezing.}
		(\textbf{A}) LLE simulation of two multiplexed OPA modes generated by sequential on-resonance seeding of $-7^{th}$ mode and $-8^{th}$ mode. The insets show intracavity power building up inside the microresonator, seeded respectively in the $-7^{th}$ and $-8^{th}$ mode, as the pump detuning increases and then is held constant at a fixed pump detuning. By choosing which mode to seed, we deterministically select which pair of microresonator modes is amplified by the OPA. (\textbf{B}) Optical spectra of the OPA input and output while pumping at the 1587 nm mode and seeding at 1561 nm mode ($\pm7^{th}$ mode). A qualitatively similar optical spectrum was measured for a \qty{1557}{\nm} seed ($\pm8^{th}$ mode). The normalization of the optical spectra in B and C are such that \qty{0}{\dB} corresponds to \qty{1}{\mW} of power. (\textbf{C}) Comparison between shot noise and intensity-difference quantum squeezed noise, demonstrating strong nanophotonic squeezing of 5.6\,dB\,$\pm$\,0.2\,dB. The error is approximated mostly with one standard deviation of the squeezed noise trace (See Supplementary Section 7). The normalization of the noise power is such that \qty{0}{\dB} corresponds to \qty{1}{\mW} of power.}
	\label{fig2} 
\end{figure}

Different from OPOs which naturally oscillate favoring the mode with the highest FWM gain (Fig.~\ref{fig1}A), our method utilizes a seed to selectively excite individual pair-wise OPA modes within the FWM gain bandwidth (Fig.~\ref{fig1}C). In the absence of a seed, these mode pairs exhibit vacuum squeezing that has been demonstrated in nanophotonics using phase-coherent local oscillators, often created through electro-optic modulation or phase-locked loops \cite{Vaidya:20, Zhang:21, Jahanbozorgi:23, Yang:21}. By injecting a weak seed signal, we stimulate regenerative amplification in the targeted mode pair - effectively converting vacuum-squeezed modes into bright, detectable beams (Fig.~\ref{fig1}C) while still remaining below the OPO threshold\,\cite{zhao2023large}. To elucidate the mechanism of seed-assisted OPA for generating and detecting quantum squeezed states, we first simulate classical frequency-multiplexed OPAs using the Lugiato-Lefever equation (LLE)\,\cite{lugiato1987spatial, moille2019pylle}. In our simulation, a strong continuous-wave pump is combined with a weak seed probe. By injecting a low-power idler seed at the $-8^{th}$ mode along with a strong pump, the seed undergoes regenerative amplification and produces a pair of bright signal and idler. As the pump is tuned further into resonance, the intracavity power of the signal and idler increase exponentially (insets in Fig.\,\ref{fig2}A). Similarly, seeding at the $-7^{th}$ idler mode not only amplifies itself but also generates a correlated signal at the $+7^{th}$ mode, while concurrently suppressing the amplification of the vacuum fluctuations in other modes and keeping them below detectable powers. 

To probe the quantum correlations between pairwise OPA modes, we perform two-mode intensity-difference squeezing measurements. We use a strongly overcoupled, foundry-fabricated \,\SiN\,microresonator with a radius of \um{50}. In the simplified experimental setup (Fig.\,\ref{fig1}C), an amplified free-running laser at 1587 nm serves as the pump, delivering an optical power around \qty{300}{\mW}, which is slightly below the OPO threshold power of \qty{310}{\mW}. An attenuated tunable laser is co-coupled with the pump to provide the weak seed signal that extracts the parametric gain to the desired OPA mode, deterministically generating the bright twin beams, as shown by the measured optical spectra in Fig.\,\ref{fig2}B. Then, we collect the light out of the chip using an antireflection-coated aspheric lens and spatially separate the pump, signal, and idler beams using a transmission grating. After diverting 1.5\,\% of each beam for power monitoring, the remaining signal and idler beams are directed to a balanced photodetector (BPD) (see supplementary information for detailed setup). We measure two-mode intensity-difference noise of 5.6~dB $\pm$ 0.2~dB below the shot-noise level with dark noise subtraction (Fig.\,\ref{fig2}C), confirming the strong quantum correlation between the generated signal and idler at $\pm7^{th}$ mode in the below-threshold OPA. The squeezing level is consistent with the prediction in a cavity OPA after accounting for the microresonator’s overcoupling efficiency and the setup’s detection loss. 

The expected squeezing level in a cavity OPA, in the ideal case, can be approximated by $S = 1-\eta_c \eta_{\text{path}} \eta_{D}$ when the pump power is near the device’s oscillation threshold at low measurement frequency\,\cite{Dutt:15,Fabre:89,Chembo:16,Vahlbruch:16}. $\eta_c$ is the overcoupling coefficient of the microresonator, which primarily depends on the ring-bus coupling gap. For a fixed gap, $\eta_c$ increases at a longer wavelength. The squeezing level is thus limited by $\eta_c \approx 93\%$ for the lower-wavelength idler at 1561~nm. The detection path efficiency has been optimized and measured to be $\eta_{\text{path}}\approx84\,\%$, while the detector efficiency of the BPD is $\eta_{D}\approx99\,\%$. The measured squeezing level is consistent with the expected value based on the overcoupling $\eta_c$ (generated on-chip squeezing $\approx$ 9~dB) after accounting for degradation from these independently characterized losses (See Supplementary Materials).

\begin{figure*}
	\centering
	\includegraphics[width=.9\textwidth]{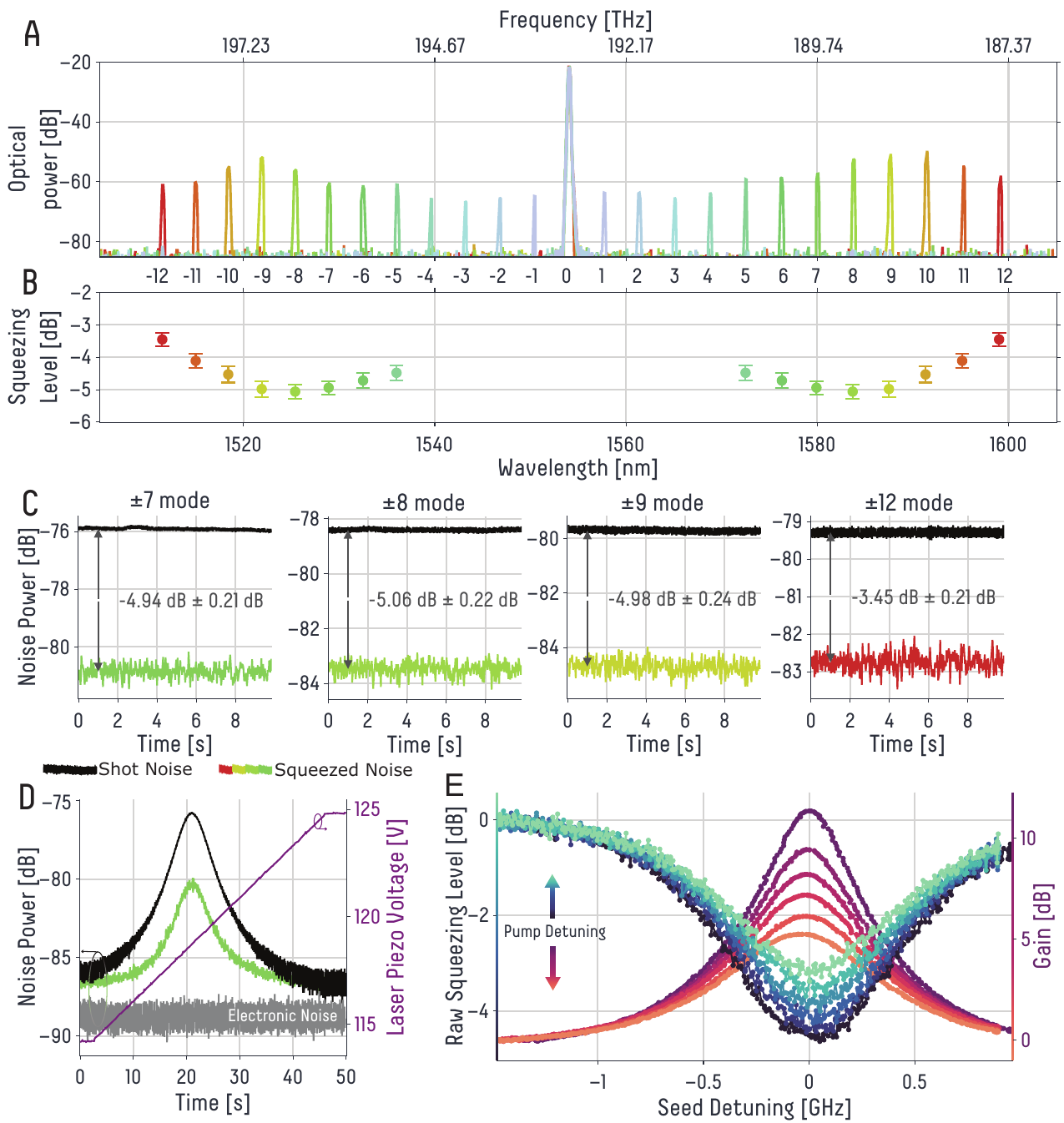} 
	\caption{\textbf{Bright squeezed quantum microcomb.}
		(\textbf{A}) Measured optical spectrum of a frequency comb generated by sequentially placing a \qty{8}{\uW} seed on resonance for each mode while pumping at 1554~nm. (\textbf{B}) Measured quantum squeezed noises for 8 mode pairs across the comb. (\textbf{C}) Representative 10-second traces of intensity-difference noise (colored) and the corresponding shot noise (black) for the $\pm7^{th}$, $\pm8^{th}$, $\pm9^{th}$, and $\pm12^{th}$ modes, measured at a 5~MHz analysis frequency on an electrical spectrum analyzer. (\textbf{D}) Intensity-difference noise (green) of the $\pm7^{th}$ mode as the seed laser’s piezo voltage (purple) changes. (\textbf{E}) Raw squeezing (without dark noise subtraction) against seed detuning for six pump detunings. Each pump detuning is separated by around 0.7 GHz. The corresponding parametric gain is shown with red shaded curves on the right axis. Lighter shades indicate larger detunings and hence smaller intracavity pump powers. The normalization is such that 0 dB corresponds to 1 mW of optical power.}
	\label{fig3} 
\end{figure*}
\section*{Bright squeezed quantum frequency comb}
To demonstrate the scalability in bandwidth and number of modes of our OPA squeezer, we experimentally show a bright amplitude-squeezed frequency comb in the same Si$_3$N$_4$ microresonator device. The microresonator is pumped with a power of around \qty{200}{\mW} at \qty{1554}{\nm} - lower than at 1587 nm pump mode due to a smaller, wavelength-dependent overcoupling coefficient. By seeding each OPA mode individually with a fixed power of \qty{8}{\uW} at a wavelength longer than the pump, we generate a multiplexed frequency comb consisting of 24 bright modes (Fig. 3A). It is important to distinguish this from a microresonator Kerr frequency comb operated far above the oscillation threshold, where soliton states, modulational instability, Turing rolls, and chaotic states can form where a large number of modes simultaneously oscillate \cite{Chembo:16,herr2014temporal,herr:16,godey:14,yang:24}. Note that the phrase "quantum frequency comb" is cursorily used by the field of quantum optics to refer to the equally spaced modes separated by the FSR even if phase coherence or mode locking between these modes is not manifest\,\cite{reimer_generation_2016,kues:19,zhu:21,Roslund:14,lu:19,Chembo:16,Yang:21,jaramillo:17,Lustig:24}, which is different from the case of a classical frequency comb. Instead, the “comb” is experimentally constructed by superimposing optical spectra obtained by sequentially probing successive resonant modes with a tunable seed. This optical spectrum, which is consistent with LLE simulations (see Supplementary Materials), characterizes parametric gain in the below-threshold resonant OPA.

We perform intensity-difference noise measurements on 8 squeezed mode pairs (Fig. 3B and 3C), achieving squeezing levels ranging from 3.5~dB $\pm$ 0.2~dB for the $\pm12^{th}$ mode with two-mode separation of 11~THz, up to 5.1~dB $\pm$ 0.2~dB for the $\pm8^{th}$ mode (Fig.\,\ref{fig3}C). The seed power at different OPA modes is adjusted by a variable optical attenuator (VOA) such that the power of each pair of generated bright twin beams provides sufficient dark noise clearance while being smaller than the BPD saturation power. Time traces of the squeezing measurement for each mode in Fig.\,\ref{fig3}B are presented in the Supplementary Materials. The seeding process brings the modes pair-by-pair above the BPD's noise floor, enabling sequential detection. The measurements in Fig.\,\ref{fig2}C reveal a two-mode squeezed vacuum generated by the OPA in the absence of seeding with the form $|{\rm TMSS}\rangle_m \propto \sum_n \tanh^n{r} |n\rangle_{s,+m} |n\rangle_{i,-m}$, where the signal $s,+m$ mode and the idler $i, -m$ mode exhibit strong quantum correlations with $r=1$. On the other hand, the measurements here in Fig. \ref{fig3}B hint at the presence of a tensor product structure $|{\rm comb}\rangle \propto \otimes_m \left[ \sum_m\tanh^n{r_m} |n\rangle_{s,+m} |n\rangle_{i,-m}\right]$, where each independent signal-idler pair is in a two-mode squeezed vacuum. 

To evaluate the bandwidth of each individual squeezed mode, we linearly sweep the frequency of the seed laser by driving the piezo actuator and measuring the squeezed noise. Figure \ref{fig3}D shows a 50-second trace of quantum noise measurement of the $\pm7^{th}$ mode alongside the piezo voltages. By calibrating the frequency change using a wavemeter, we extract the raw, or without dark noise subtracted, squeezed noise trace as a function of seed laser detuning. The maximum squeezing is observed at the zero detuning point, where the seed experiences the largest parametric gain. We then proceed to measure the squeezed noises at six pump detunings spanning about \qty{3.5}{\GHz} by varying the currents on the integrated microheaters. As the pump laser shifts away from pump resonance, the maximum gain at on-resonance seed frequency decreases (Fig.\,\ref{fig3}E, dark to light red), lowering the raw squeezing level due to the reduced twin-beam powers. However, the dark-noise-subtracted squeezing levels remain largely unaffected. (see Supplementry Materials). 

\begin{figure*} 
	\centering
	\includegraphics[width=.9\textwidth]{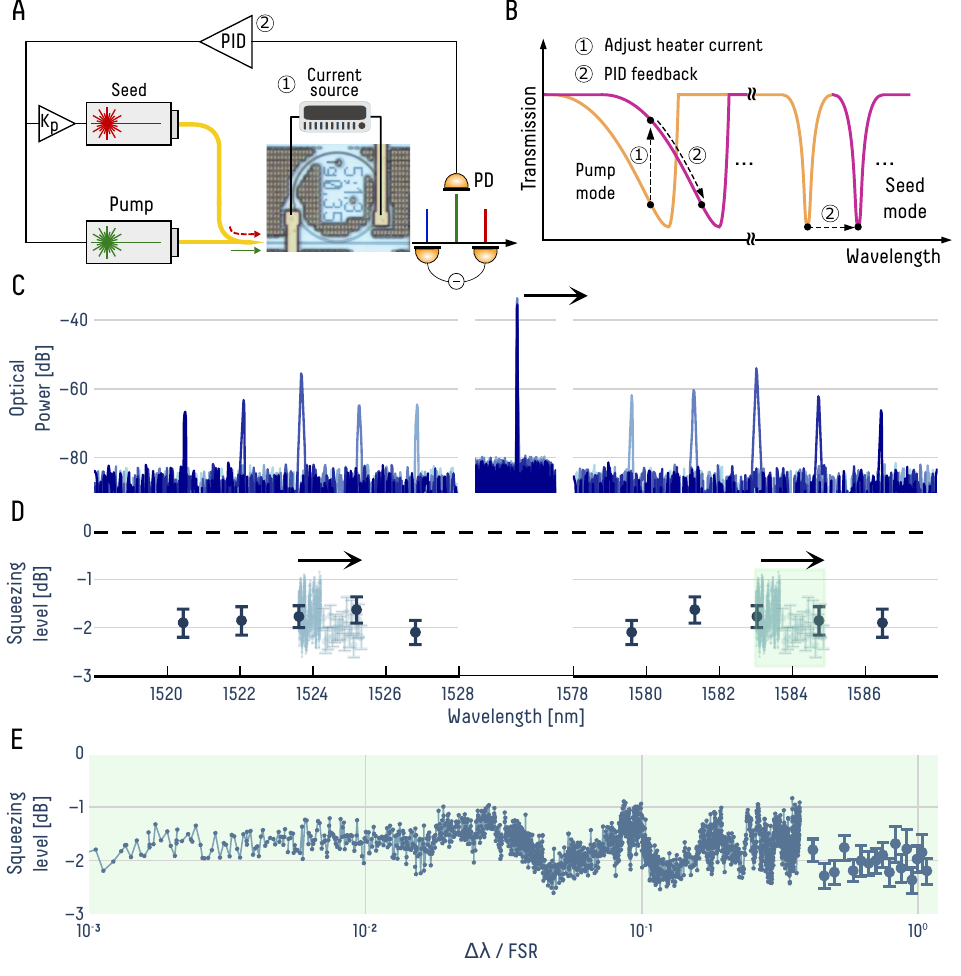} 
	\caption{\textbf{Continuous spectral tunability of squeezed light source.}
		(\textbf{A}) Schematic diagram showing the resonance control and electronic feedback for continuous tuning of the pump and seed lasers. (\textbf{B}) Schematic of spectral tuning mechanism. (\textbf{C}) Measured optical spectra of five individually seeded two-mode pairs and (\textbf{D}) the  associated squeezing levels. The normalization is such that 0 dB corresponds to 1 mW of optical power. (\textbf{E}) Fine continuous and coarse discrete tuning achieved by actuating the microheater atop the ring resonator. The tuning covers more than one FSR.
}
	\label{fig4} 
\end{figure*}

\section*{Spectral Tuning of squeezed mode}
We experimentally demonstrate continuous wavelength tunability of a squeezed qumode pair spanning one FSR of the microresonator, bridging the gap between discrete comb lines. To demonstrate the potential of \SiN platform in on-chip quantum-enhanced sensing, we choose a different microresonator with transverse magnetic mode excitation and a FSR of 210 GHz. The measured optical spectrum and squeezing level of 10 qumodes are shown in Fig.~\ref{fig4}C and~\ref{fig4}D. The reduced squeezing level here compared to the result in Fig.\,\ref{fig3} is due to lower overcoupling ($\approx 65~\%$) of the microresonator and lower BPD efficiency ($\approx 79~\%$) (see Supplementary Materials). 

To fill the spectral gap between two adjacent squeezed modes, we use a proportional-integral-derivative (PID) controller to drive the on-chip microheater, pump laser, and seed laser coherently. We first pump the microresonator at 1554 nm and seed at 1583 nm ($+17^{th}$ mode). The on-chip microheater is then driven via a DC current source, which shifts all the microresonator resonances to longer wavelengths due to the thermo-optic effect (purple curve in Fig.~\ref{fig4}B). This shift detunes the seed and pump laser frequencies from the resonances, manifesting as a jump in the transmitted pump power (Fig. \ref{fig4}A). The error signal is defined as the difference between the transmitted pump power and a constant setpoint. A PID compensator applied on the pump laser piezo allows it to follow the detuning of the resontaor. To ensure the frequency shift of the seed laser tracks the pump laser by the same amount, an appropriately scaled PID signal is also sent to the seed (Fig. \ref{fig4}A, indicated by $K_{P}$). This approach is effective within the frequency range of a few gigahertz when the piezo response of the pump and seed lasers are linear.

Using the PID control scheme, we demonstrate continuous fine-tuning with a step of \qty{0.2}{\pm} in wavelength, which is three-orders of magnitude smaller than the FSR, across the first \qty{0.63}{\nm} range (connected points in Fig.~\ref{fig4}E). In the remaining spectral range within one FSR, we also show spectral shift of the squeezed modes without active feedback control, at the cost of scanning speed and frequency resolution (discrete points in Fig.~\ref{fig4}E). The squeezing levels maintain between 1~dB and 2.3~dB in this one-FSR span, indicating the agility and robustness of seed-assisted OPA in spectral qumode tuning. 

\begin{figure*}
	\centering
	\includegraphics[width=.9\textwidth]{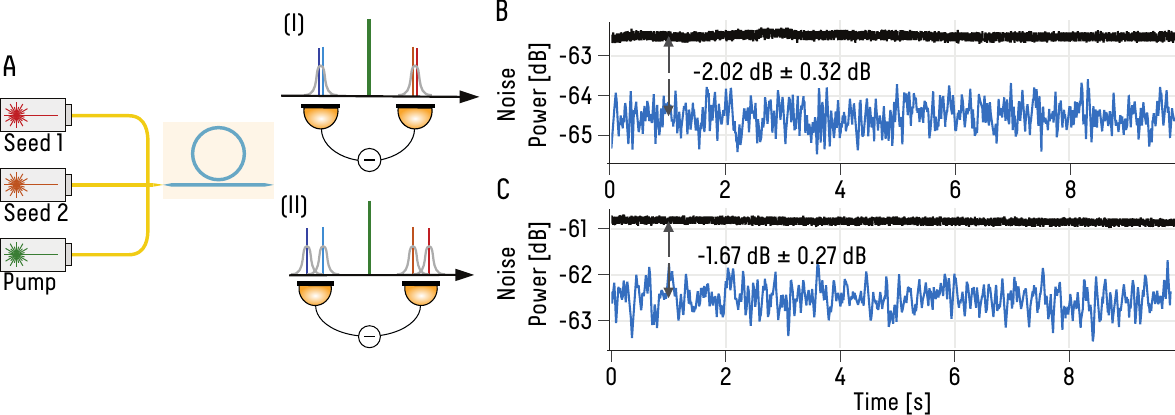} 
	\caption{\textbf{Multiseeded quantum squeezing.}
		(\textbf{A}) Two uncorrelated lasers, serving as dual-color seed signals, are simultaneously amplified by the OPA. Two distinct two-mode pairs are detected simultaneously at the balanced detector for I: same mode and II: adjacent modes. Shot noise and quantum squeezed noises from the dual two-mode squeezing measurements for (\textbf{B}) two seeds in the same resonance mode and (\textbf{C}) two seeds in two resonance modes. The normalization is such that 0 dB corresponds to 1 mW of noise power.}
	\label{fig5} 
\end{figure*}

\section*{Multiseeded Squeezing}
Finally, to demonstrate the scalability of our system in both bandwidth and the number of squeezed modes, we performed a multiseeded experiment where two seed lasers were simultaneously sent to the OPA. Figure \ref{fig5}A shows two different seeding configurations. In the first configuration, two uncorrelated seed lasers were independently tuned into the same resonant mode of a Si$_3$N$_4$ microresonator ($m=17$) with 210 GHz FSR, alongside the strong continuous-wave pump. In the second configuration, two seeds were tuned into two adjacent modes ($m=17 ~\text{and}~ 18$) of the microresonator.

To detect the quantum correlations, the output was routed such that the two signal beams were directed to one input port of the BPD, and the two idler beams to the other. We measured an intensity-difference noise of 2.0~dB below the shot noise level in the first configuration and 1.7~dB in the second configuration. The quantum noise reductions are comparable to the single-pair squeezing levels that we observed for the same device in Fig. \ref{fig4}. These results confirm that our seed-assisted OPA technique can be expanded to measure simultaneous bright squeezing of multiple independent mode pairs, paving the way for frequency-multiplexed continuous-variable quantum processing using integrated microresonators. 

\section*{Discussion}
In summary, we introduced the seed-assisted OPA approach into nanophotonics to realize a squeezed microcomb featuring high two-mode-squeezing, broad spectral coverage, and agile frequency tuning. The maximum observed squeezing is primarily constrained by $\approx\,84\,\%$ of detection path efficiency (which includes outcoupling efficiency from the chip of $\approx\,92\,\%$), and the microresonator's overcoupling coefficient $\eta_c\approx93\,\%$. As we expected and observed, the lower-wavelength idler resonance usually has a smaller $\eta_c$, setting the maximum squeezing level, whereas the power threshold is determined by the higher $\eta_c$ of the pump resonance. A possible approach to get higher squeezing is to design a device in which $\eta_c$ of the idler and signal resonances are nearly unity, while also ensuring a manageable pump power by using a higher-quality-factor \cite{Ji:17,puckett:21}, a smaller FSR ring \cite{shen:25}, or a lower $\eta_c$ for the pump.

Looking ahead, expanding both the bandwidth and the number of squeezed modes can be pursued by tailoring the microresonator’s dispersion and FSR. In our current setup, the 11-THz separation of the widest squeezed qumode pair is set by the available laser wavelength range (1510 to 1630 nm), although the available FWM gain window should span a wider range. Wider frequency spacings can be accessed by engineering a broader FWM gain window, whereas designing a smaller-FSR resonator would increase the number of qumodes in a single device, albeit at the cost of higher pump power threshold. Furthermore, the fine and coarse tunability we have demonstrated, in particular for the vertically delocalized TM waveguide mode, highlights the potential of these devices for cladding-integrated evanescent spectroscopic sensing. In such scenarios, frequency-agile quantum correlations can be leveraged to enhance measurement sensitivities across multiple spectral channels.

The ability to generate and measure strong, frequency-multiplexed squeezing in a compact device underpins the promise of integrated continuous-variable architectures for quantum sensing and information processing. In QIP, it has been typically assumed that below-threshold vacuum squeezers are preferred over above-threshold ones. Consistent with this assumption, our below-threshold quantum microcomb constitutes the critical quantum resource, with seed-assisted measurement enabling characterization of the high-quality nature of the squeezed microcomb consisting of frequency-multiplexed two-mode squeezed vacua exhibiting large levels of quantum noise reduction. On the other hand, quantum sensing can benefit significantly from both above-threshold \cite{herman:25,Samantaray:17,pooser_ultrasensitive_2015,adamou:25} and below-threshold \cite{Vahlbruch:16, acernese:19, lucivero:16, wolfgramm:10} squeezers. By further combining on-chip control elements (tunable filters, phase shifters, and on-chip balanced detectors \cite{Nie:19,Singh:22,Gao:23,Costanzo:21,Bruynsteen:21,Raffaelli:18,Tasker:21,Gurses:24,Tasker:24}), one can envision the creation of fully integrated quantum photonic platforms, and our seed-assisted microresonator OPA serves as a practical architecture towards quantum advantage in this direction.

\begin{acknowledgments}
The authors acknowledge fruitful scientific discussions with Zhifan Zhou and Paul Lett. We thank Samyak Gothi and David Long for their insightful feedback. We acknowledge Synopsys Optical Solutions for providing simulation and chip layout software. Mention of specific companies or trade names is for scientific communication only, and does not constitute an endorsement by NIST. This work was funded by grants from the National Science Foundation (QuSeC-TAQS \# 2326792, CAREER \# 2340835) and NAWCAD (\# N004212310002).
\end{acknowledgments}

\appendix
\section*{Methods}
\subsection*{Device parameters}
Our microring resonators were fabricated through the multi-project-wafer (MPW) service. The film thickness of the buried oxide, \SiN waveguide, and top SiO\textsubscript{2} cladding are \qty{4}{\um}, \qty{780}{\nm}, and \qty{6.6}{\um}, respectively. Table~\ref{table1} summarizes the design parameters of the two microresonators used in Fig. \ref{fig2}, \ref{fig3} and Fig. \ref{fig4}, \ref{fig5}.

\begin{table}[h]
	\centering
	\caption{\textbf{Summary of Si$_3$N$_4$ microring device parameters.}}
	\label{table1}
	\begin{tabular}{lcc} 
		\hline
		Parameter & Fig.\ 2, 3 & Fig.\ 4, 5 \\
		\hline
		Radius [\um{}] & 50  & 110 \\
		FSR [GHz] & 450 & 210 \\
		Gap [\um{}] & 0.3 & 0.3 \\
		Width [\um{}] & 1.3 & 1.7 \\
		Pol. mode & $TE_0$ & $TM_0$ \\
		$Q_i$ [million] & $\approx 1.5$ & $\approx 2.0$ \\
		$Q_L$ [million] & $\approx 0.1$ & $\approx 0.5$ \\
		Measured $\beta_2$ [fs$^2$/mm] & -297.7 & to do\\
		\hline
	\end{tabular}
\end{table}

\subsection*{Lasers and balanced detectors}
Table \ref{table3} and \ref{table2} summarize the BPDs and lasers used in the experiments. Table~\ref{table4} records the configuration of pump laser, seed laser, BPD, and corresponding figures.
\begin{table}
	\centering
	\caption{\textbf{Summary of BPDs used in the experiments.} QE: quantum efficiency.}
	\label{table3}
	\begin{tabular}{lccc} 
		\hline
		Name & QE & RF Bandwidth [MHz] & Monitor output\\
		\hline
		BPD-1 & 79\,\% & 75 & Yes\\
		BPD-2 & 88\,\% & 75 & Yes \\
        BPD-3 & 99\,\% & 80 & No \\
		\hline
    \end{tabular}
\end{table}
\begin{table*}
	\centering
	\caption{\textbf{Summary of lasers used in the experiment}}
	\label{table2}
	\begin{tabular}{lccc} 
		\hline
		Device name & Wavelength [nm] & Linewidth (\qty{5}{\us}) [kHz] & Piezo-tuning range [GHz] \\
		\hline
		Tunable Laser A & $1510 - 1630$ & $\leq$10 & 33\\
		Tunable Laser B & $1520 - 1570$  & $\leq$100 & 30 \\
        Single-wavelength Laser C & 1554.13 & $\leq$1 & 0.2 \\
		\hline
    \end{tabular}
\end{table*}

\begin{table}[h] 
	\centering
	\caption{\textbf{Summary of laser and BPD configurations used in each figure.}}
	\label{table4}
	\begin{tabular}{lccc} 
		\hline
		Figure & Pump Laser & Seed Laser & BPD \\
		\hline
		Fig. 2 & Laser A & Laser B & BPD-3 \\
		Fig. 3 & Laser C  & Laser A & BPD-3\\
		Fig. 4 & Laser C  & Laser A & BPD-1 \\
		Fig. 5 & Laser C  & Laser A \& B & BPD-1 \\
		Fig. S3 & Laser C & Laser A & BPD-2 \\
		\hline
	\end{tabular}
\end{table}
\subsection*{Extended data for squeezed microcomb}
\begin{figure*}[h]
	\centering
	\includegraphics[width = 1\textwidth]{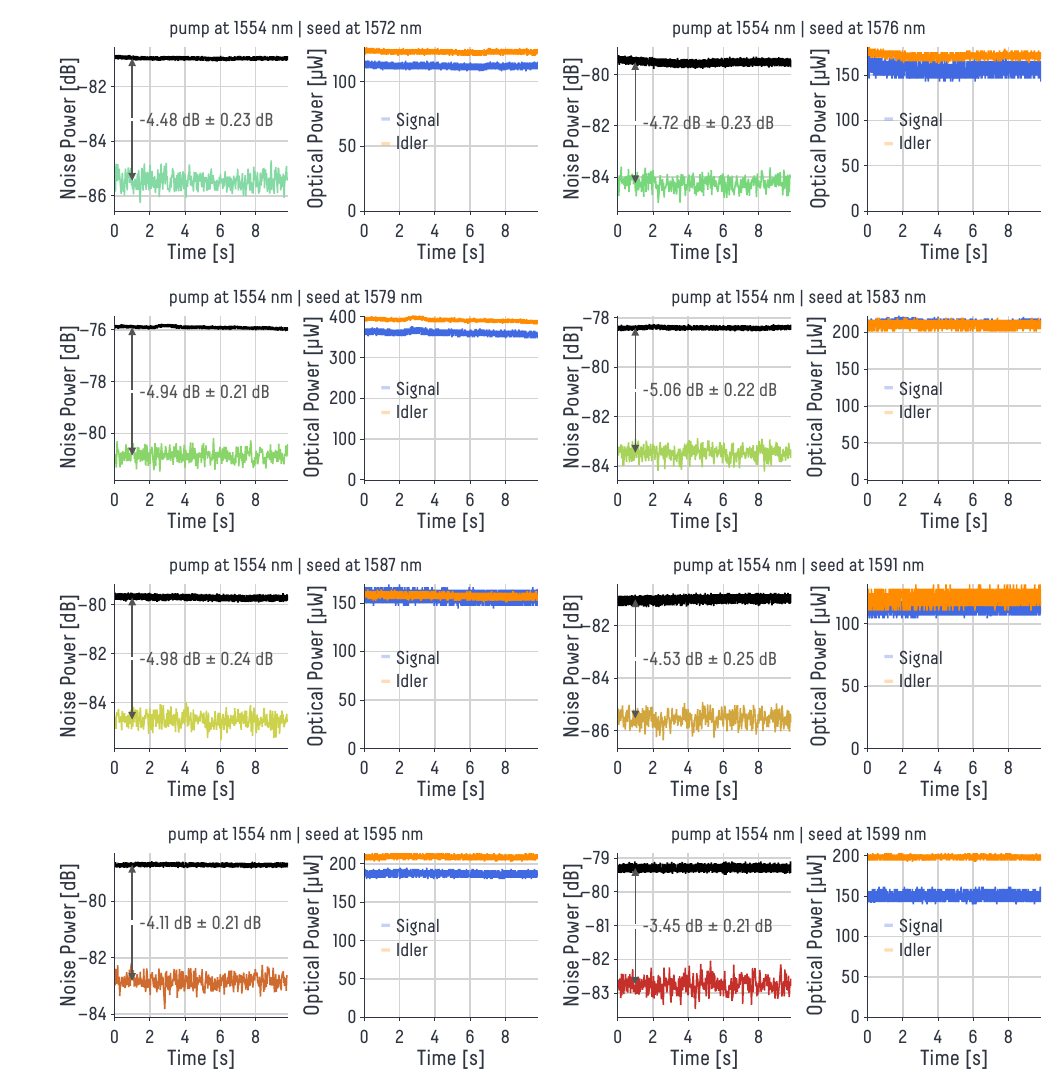}
	\caption{For each pairsise squeezed comb line, (left) a comparison of squeezed noises and shot noises, (right) the corresponding optical powers for signal (orange) and idler (blue). The normalization is such that 0 dB corresponds to 1 mW of noise power.}
	\label{figSqCombExtent} 
\end{figure*}

We present the squeezing measurements for the 16 squeezed comb modes shown in Fig.\,\ref{fig3}. For each qumode, the left panel of Fig.\,\ref{figSqCombExtent} compares the measured squeezed noise and the shot noise, while the right panel plots the corresponding twin-beam optical powers. Since we are seeding the higher-wavelength signal, its power is higher than the idler field. For the modes with higher FWM gain, such as $\pm 8^{th} $ (\qty{1583}{\nm}) mode and $\pm 9^{th} $ (\qty{1587}{\nm} mode), the signal and idler powers are nearly identical, reflecting lower seed power being used. As the FWM gain decreases for farther apart modes from the pump, a larger-power seed is needed, which manifests as a more noticable power difference. 

\bibliography{apssamp}

\end{document}